\documentclass[12pt]{article}
\usepackage{graphicx}

\def\hybrid{\topmargin 0pt      \oddsidemargin 0pt
        \headheight 0pt \headsep 0pt
       \voffset-1cm
        \textwidth 6.25in       
       \textheight 9.5in       
        \marginparwidth 0.0in
        \parskip 5pt plus 1pt   \jot = 1.5ex}
\catcode`\@=11
\def\marginnote#1{}

\newcount\hour
\newcount\minute
\newtoks\amorpm
\hour=\time\divide\hour by60
\minute=\time{\multiply\hour by60 \global\advance\minute by-\hour}
\edef\standardtime{{\ifnum\hour<12 \global\amorpm={am}%
        \else\global\amorpm={pm}\advance\hour by-12 \fi
        \ifnum\hour=0 \hour=12 \fi
        \number\hour:\ifnum\minute<10 0\fi\number\minute\the\amorpm}}
\edef\militarytime{\number\hour:\ifnum\minute<10 0\fi\number\minute}

\def\draftlabel#1{{\@bsphack\if@filesw {\let\thepage\relax
   \xdef\@gtempa{\write\@auxout{\string
      \newlabel{#1}{{\@currentlabel}{\thepage}}}}}\@gtempa
   \if@nobreak \ifvmode\nobreak\fi\fi\fi\@esphack}
        \gdef\@eqnlabel{#1}}
\def\@eqnlabel{}
\def\@vacuum{}
\def\draftmarginnote#1{\marginpar{\raggedright\scriptsize\tt#1}}

\def\draftlabel#1{{\@bsphack\if@filesw {\let\thepage\relax
   \xdef\@gtempa{\write\@auxout{\string
      \newlabel{#1}{{\@currentlabel}{\thepage}}}}}\@gtempa
   \if@nobreak \ifvmode\nobreak\fi\fi\fi\@esphack}
        \gdef\@eqnlabel{#1}}
\def\@eqnlabel{}
\def\@vacuum{}
\def\draftmarginnote#1{\marginpar{\raggedright\scriptsize\tt#1}}

\def\draft{\oddsidemargin -.5truein
        \def\@oddfoot{\sl preliminary draft \hfil
        \rm\thepage\hfil\sl\today\quad\militarytime}
        \let\@evenfoot\@oddfoot \overfullrule 3pt
        \let\label=\draftlabel
        \let\marginnote=\draftmarginnote
   \def\@eqnnum{(\theequation)\rlap{\kern\marginparsep\tt\@eqnlabel}%
\global\let\@eqnlabel\@vacuum}  }

\def\numberbysection{\@addtoreset{equation}{section}
        \def\theequation{\thesection.\arabic{equation}}}

\def\underline#1{\relax\ifmmode\@@underline#1\else
        $\@@underline{\hbox{#1}}$\relax\fi}

\def\titlepage{\@restonecolfalse\if@twocolumn\@restonecoltrue\onecolumn
     \else \newpage \fi \thispagestyle{empty}\c@page\z@
        \def\thefootnote{\fnsymbol{footnote}} }

\def\endtitlepage{\if@restonecol\twocolumn \else  \fi
        \def\thefootnote{\arabic{footnote}}
        \setcounter{footnote}{0}}  
\relax

\hybrid

\newfont{\Bbb}{msbm10 scaled 1\@ptsize00}
\newfont{\Bbbb}{msbm7 scaled 1\@ptsize00}

\newcommand{\DDD}{\raise-1pt\hbox{$\mbox{\Bbbb D}$}}

\newcommand{\UUU}{\raise-1pt\hbox{$\mbox{\Bbbb U}$}}

\newcommand{\z}{\raise-1pt\hbox{$\mbox{\Bbbb Z}$}}

\def\beq{\begin{equation}}
\def\eeq{\end{equation}}
\def\p{\partial}

\begin{document}

\begin{titlepage}

\title{Dispersionless Pfaff-Toda hierarchy
and elliptic L\"owner equation}

\author{V.~Akhmedova\thanks{
International Laboratory of Representation 
Theory and Mathematical Physics,
National Research University Higher School of Economics,  
20 Myasnitskaya Ulitsa, Moscow 101000, Russia,
e-mail: valeria-58@yandex.ru}
\and A.~Zabrodin
\thanks{Institute of Biochemical Physics,
4 Kosygina st., Moscow 119334, Russia; ITEP, 25
B.Cheremushkinskaya, Moscow 117218, Russia and
International Laboratory of Representation 
Theory and Mathematical Physics,
National Research University Higher School of Economics,
20 Myasnitskaya Ulitsa,
Moscow 101000, Russia, e-mail: zabrodin@itep.ru}}

\date{May 2016}
\maketitle

\vspace{-7cm} \centerline{ \hfill ITEP-TH-07/16}\vspace{7cm}

\begin{abstract}

We show that one-variable reductions of the Pfaff-Toda integrable
hierarchy in the dispersionless limit are described by a system of coupled 
elliptic L\"owner (Komatu-Goluzin) equations.

\end{abstract}

\end{titlepage}

\vspace{5mm}

%



\section{Introduction}

In this paper we consider the dispersionless limit
of the Pfaff-Toda hierarchy \cite{Willox,Takasaki09}
in the elliptic parametrization \cite{AZ14,AZ15}.
The aim of the paper is to describe one-variable 
reductions of the hierarchy in the spirit of the 
Gibbons-Tsarev approach \cite{GT1,GT2}, see also
\cite{Manas1}-\cite{Tak2014}, where it was shown that 
consistent reductions of the KP, modified KP and Toda hierarchies are 
obtained from solutions to the chordal and radial versions of the L\"owner equation 
known in the classical complex analysis \cite{Pommerenke}. 

A similar description of 
reductions of the Pfaff-KP hierarchy (also known as 
the Pfaff lattice) is given in \cite{AZ14}, where
it has been shown that in the Pfaff-KP case the L\"owner equation 
should be substituted by its elliptic analogue (the Komatu-Goluzin 
equation \cite{Komatu,Goluzin}, see also \cite{Alexandrov}-\cite{Zhan}).
Here we show that in the Pfaff-Toda case the reductions are described by solutions to
a system of coupled equations looking like 
the elliptic L\"owner ones. The proof is based on some non-trivial identities
for theta functions.

\section{The dispersionless Pfaff-Toda hierarchy}

The set of hierarchical times in the dispersionless Pfaff-Toda (dPfaff-Toda) hierarchy
is ${\bf t}=
\{\ldots , \bar t_2, \bar t_1, \bar t_0,
t_0, t_1, t_2, \ldots \}$. Below we deal with the real form 
of the hierarchy, where $\bar t_k$ is the complex conjugate of $t_k$ for all $k=0, 1, 2, 
\ldots$.
Let us introduce the differential operators
\beq\label{DD1}
D(z)=\sum_{k\geq 1}\frac{z^{-k}}{k}\, \p_{t_k}, \qquad \phantom{a}
\bar D(z)=\sum_{k\geq 1}\frac{z^{-k}}{k}\, \p_{\bar t_k}
\eeq
and
\beq\label{DD2}
\nabla (z)=\p_{t_0} +D(z), \qquad \phantom{a}
\bar \nabla (z)=\p_{\bar t_0} +\bar D(z).
\eeq
Clearly, $\overline{D(z)}=\bar D(\bar z)$, $\overline{\nabla (z)}=
\bar \nabla (\bar z)$. In the dispersionless 
Hirota formulation, 
the dependent variable of the dPfaff-Toda hierarchy is a real function
$F=F({\bf t})$. 
Introducing the auxiliary functions
\beq\label{T1}
\begin{array}{l}
P(z)=ze^{-(\p_{t_0}+\p_{\bar t_0})\nabla (z)F}, \qquad \phantom{a}
W(z)=ze^{-(\p_{t_0}-\p_{\bar t_0})\nabla (z)F},
\\ \\
\bar P(z)=ze^{-(\p_{t_0}+\p_{\bar t_0})\bar \nabla (z)F}, \qquad \phantom{a}
\bar W(z)=ze^{(\p_{t_0}-\p_{\bar t_0})\bar \nabla (z)F},
\end{array}
\eeq
one can present equations of the hierarchy in the form \cite{Takasaki09}
\beq\label{T2}
\begin{array}{l}
\displaystyle{
e^{D(z)D(\zeta )F}\left (1-\frac{1}{P(z)P(\zeta )}\right )=\phantom{}
\frac{W(z)-W(\zeta )}{z-\zeta}\, e^{(\p_{t_0}-\p_{\bar t_0})\p_{t_0}F}}\phantom{}
\\ \\
\displaystyle{
e^{D(z)D(\zeta )F}\left (1-\frac{1}{W(z)W(\zeta )}\right )=\phantom{}
\frac{P(z)-P(\zeta )}{z-\zeta}\, e^{(\p_{t_0}+\p_{\bar t_0})\p_{t_0}F}}\phantom{}
\\ \\
\displaystyle{
e^{D(z)\bar D(\bar \zeta )F}\left (1-\frac{1}{P(z)\phantom{}
\overline{P(\zeta)}}\right )=\phantom{}
1-\frac{1}{W(z)\overline{W(\zeta)}}}\phantom{}
\\ \\
\displaystyle{
e^{D(z)\bar D(\bar \zeta)F}\left (W(z)-\overline{W(\zeta)}\right )=\phantom{}
\left (P(z)-\overline{P(\zeta)}\right )\phantom{}
e^{2\p_{t_0}\p_{\bar t_0} F }}.
\end{array}
\eeq
Here $\overline{P(\zeta)}:=\bar P(\bar z)$,
$\overline{W(\zeta)}:=\bar W(\bar z)$.
There are also two equations which are complex conjugate to 
the first and the second equation in the list. (The third and 
the fourth ones are self-conjugate.)
The differential equations of the Pfaff-Toda 
hierarchy are obtained by expanding 
(\ref{T2}) in powers of $z$ and $\zeta$.

Dividing the second equation in (\ref{T2}) by the first one, we get the relation
$$
W(z)+W^{-1}(z)-e^{2\p_{t_0}\p_{\bar t_0}F}
\! \left (P(z)+P^{-1}(z)\right )=\phantom{}
W(\zeta )+W^{-1}(\zeta)-
e^{2\p_{t_0}\p_{\bar t_0}F}
\! \left (P(\zeta)+P^{-1}(\zeta)\right ).
$$
It means that 
$C:= W(z)+W^{-1}(z)-e^{2 \p_{t_0}\p_{\bar t_0}F}\! 
\left (P(z)+P^{-1}(z)\right )$ does not depend on $z$.
The constant $C$ can be found by performing the   
limit $z\to \infty$:
$
C=2e^{-(\p_{t_0}-\p_{\bar t_0})\p_{t_0}F}\p_{\bar t_0}\p_{t_1}F
$.
Dividing the fourth equation in (\ref{T2}) by the third one,
we get the relation 
$$
W(z)+W^{-1}(z)-e^{2\p_{t_0}\p_{\bar t_0}F}
\! \left (P(z)+P^{-1}(z)\right )=\phantom{}
\bar W(\bar \zeta )+\bar W^{-1}(\bar \zeta )-e^{2\p_{t_0}\p_{\bar t_0}F}
\! \left (\bar P(\bar \zeta )+\bar P^{-1}(\bar \zeta )\right )
$$
which states that $C$ is real, i.e.,
$e^{\p_{t_0}^2 F}\p_{t_0}\p_{\bar t_1}F=e^{\p_{\bar t_0}^2F}
\p_{\bar t_0}\p_{t_1}F$ (this is in fact one of equations of the hierarchy). 
As a result, the auxiliary functions $P(z), W(z)$ satisfy the algebraic
equation \cite{Takasaki09}
\beq\label{T3}
W(z)+W^{-1}(z)-R^2\left ( P(z)+P^{-1}(z)\right )=C\phantom{}
\eeq
with the real coefficients
\beq\label{T4}
R^2=e^{2\p_{t_0}\p_{\bar t_0}F}, \quad \phantom{a}
C= 2e^{\p_{t_0}(\p_{\bar t_0}-\p_{t_0})F}\p_{\bar t_0}\p_{t_1}F.
\eeq
The functions $\bar P$, $\bar W$ satisfy the same equation.
This equation
defines an elliptic curve, with $P$ and $W$ being 
functions on it. The local parameter around $\infty$ is $z^{-1}$. 
As is readily seen from (\ref{T1}), 
both $P$ and $W$ have a simple pole at infinity.

Along with $P$, $W$ consider the functions
\beq\label{T5}
f(z)=\sqrt{P(z)W(z)}=ze^{-\p_{t_0}\nabla (z)F}, \qquad \phantom{a}
g(z)=\sqrt{\frac{P(z)}{W(z)}}=e^{-\p_{\bar t_0}\nabla (z)F}.
\eeq
The function $f$ has a simple pole at $\infty$ while $g$ is regular
there.
Their complex conjugate functions are 
$\overline{f(z)}=\bar f(\bar z)=
\bar ze^{-\p_{\bar t_0}\bar \nabla (\bar z)F}$,
$\overline{g(z)}=\bar g(\bar z)=
e^{-\p_{t_0}\bar \nabla (\bar z)F}$.
In terms of $f$, $g$ the equation of the elliptic curve (\ref{T3}) reads
\beq\label{T6}
R^2(f^2g^2 +1)+Cfg=f^2+g^2.
\eeq
The functions 
$\bar f(z)$, $\bar g(z)$ obey the same equation. 
Note the symmetry $f\leftrightarrow g$. 

We uniformize the elliptic curve
(\ref{T6}) with the help of the
theta functions $\theta_a(u)=\theta_a(u|\tau )$, $a=1,2,3,4$
(see the appendix for their definition):
\beq\label{ET1}
f(z)=\frac{\theta_4(u(z))}{\theta_1(u(z))}, \quad \phantom{a}
g(z)=\frac{\theta_4(u(z)+\eta )}{\theta_1(u(z)+\eta )}.
\eeq
The equation (\ref{T6}) of the curve is then satisfied
identically if
\beq\label{ET2}
R=\frac{\theta_1(\eta)}{\theta_4(\eta)}, \quad  \phantom{a}
C=2\, \frac{\theta_4^2(0)\, 
\theta_2(\eta)\, \theta_3(\eta)}{\theta_4^2(\eta)\, 
\theta_2(0)\, \theta_3(0)}.
\eeq
The modular parameter $\tau$ of the curve (with ${\rm Im} \tau >0$) and the
parameter $\eta$ (a point on the elliptic curve) 
are dynamical variables which depend on all 
the times: $\eta = \eta ({\bf t})$, $\tau = \tau ({\bf t})$.
We assume that $\eta$ is real 
and $\tau$ is purely imaginary. This assumption is consistent with 
reality of $R$ and $C$.
We fix the expansions of the functions $u(z)=u(z, {\bf t})$, 
$\bar u (z)=\bar u(z, {\bf t})$
around $\infty$ as follows:
\beq\label{E3a}
u(z, {\bf t})=
\frac{c_1({\bf t})}{z}+\phantom{}
\frac{c_2({\bf t})}{z^2}+\ldots \, ,
\qquad \phantom{}
\bar u(z, {\bf t})=\frac{\overline{c}_1({\bf t})}{z}+\phantom{}
\frac{\overline{c}_2({\bf t})}{z^2}+\ldots \, .
\eeq 
In contrast to the dispersionless Pfaff-KP hierarchy the coefficients 
$c_i$ are complex. 

After the uniformization only two of the four equations in (\ref{T2})
remain independent. 
Following \cite{AZ15}, we represent them in the elliptic form.
Let us rewrite the first and the third equations as
$$
(z_1^{-1}-z_2^{-1})e^{\nabla_1\nabla_2 F}\phantom{}
=R^{-1}g_1g_2 \, \frac{W_1-W_2}{1-P_1P_2},
$$
$$
e^{\nabla_1\bar \nabla_2 F}\phantom{}
=R^{-1}g_1\bar g_2 \, \frac{1-W_1\bar W_2}{1-P_1\bar P_2},
$$
where $\nabla_i =\nabla (z_i)$, $\bar \nabla_i =\bar \nabla (\bar z_i)$,
$g_i=g(z_i)$, etc. The identities
$$
\frac{W_1-W_2}{1-P_1P_2}\, =\,\, \phantom{}\frac{\theta_1(\eta)}{\theta_4(\eta)}\,
\frac{\theta_1(u_1+\eta)\, \theta_1(u_2+\eta)}{\theta_4(u_1+\eta)\, \phantom{}
\theta_4(u_2+\eta)}\, \cdot \frac{\theta_1(u_1-u_2)}{\theta_4(u_1-u_2)},
$$
$$
\frac{1-W_1\bar W_2}{1-P_1\bar P_2}\, =\phantom{}
\,\, \frac{\theta_1(\eta)}{\theta_4(\eta)}\,
\frac{\theta_1(u_1+\eta)\, \theta_1(\bar u_2+\eta)}{\theta_4(u_1+\eta)\,
\theta_4(\bar u_2+\eta)}\, \cdot \phantom{}
\frac{\theta_1(u_1+\bar u_2+\eta)}{\theta_4(u_1+\bar u_2+\eta)}\phantom{}
$$
allow us to represent the equations in the form \cite{AZ15}
\beq\label{ET3}
\phantom{a}
\begin{array}{rll}
(z_1^{-1}-z_2^{-1})\, e^{\nabla (z_1)\nabla (z_2)F}&=&\displaystyle{
\phantom{}\frac{\theta_1 (u(z_1)-u(z_2))}{\theta_4 (u(z_1)-u(z_2))},}
\\ &&\\
e^{\nabla (z_1)\bar \nabla (z_2)F}&=&\displaystyle{\phantom{}
\frac{\theta_1 (u(z_1)+\bar u(z_2)+\eta )}{\theta_4 
(u(z_1)+\bar u( z_2)+\eta )},}
\\ &&\\
(z_1^{-1}-z_2^{-1})\, e^{\bar \nabla (z_1)\bar 
\nabla (z_2)F}&=&\displaystyle{\phantom{}
\frac{\theta_1 (\bar u(z_1)-
\bar u(z_2))}{\theta_4 (\bar u(z_1)-\bar u(z_2))}.}
\end{array}
\eeq
Note that the first equation (a ``half'' of the 
dPfaff-Toda hierarchy with fixed times $\bar t_k$) looks like the 
equation for the dispersionless Pfaff-KP hierarchy in the elliptic 
pa\-ra\-met\-ri\-za\-tion \cite{AZ14,AZ15} (however, with complex times $t_k$).
The third equation is conjugate of the first one.
It represents another copy of the 
dispersionless Pfaff-KP hierarchy, with respect to the times 
$\bar t_k$ with fixed $t_k$'s. The second equation 
contains mixed  
$t_k$- and $\bar t_k$-derivatives. It just
couples the two hierarchies into the bigger one.

The limit $z_2\to \infty$ of equations (\ref{ET3}) yields:
\beq\label{ET4}
e^{\p_{t_0}\nabla (z)F}=z\, \frac{\theta_1(u(z))}{\theta_4(u(z))},
\quad \phantom{a}
e^{\p_{\bar t_0}\nabla (z)F}\, =\, 
\frac{\theta_1(u(z)+\eta )}{\theta_4(u(z)+\eta )}
\eeq
(and complex conjugate equations).
These are expressions for the functions
$f$ and $g$ (\ref{ET1}) combined with their definition
(\ref{T5}). The further expansion of the first relation as $z\to \infty$
gives
\beq\label{rho}
\rho:= e^{\p^2_{t_0}F}=\pi c_1\theta_2(0)\theta_3(0),
\eeq
where we have used the identity (\ref{theta1prime}) from the appendix.
This quantity is used for calculations in the next section.

It is convenient to introduce the function 
\beq\label{SSS}
S(u)=S(u|\tau ):=\log \frac{\theta_1(u|\tau )}{\theta_4 (u|\tau )}
\eeq
(up to a constant, it is logarithm of the elliptic sinus function ${\rm sn}$).
It has the quasi-periodicity properties
$S(u+1)=S(u)+i\pi$, $S(u+\tau )=S(u)$.
Its $u$-derivative is given by
$$
S'(u)=\p_u S(u|\tau )=\pi  \theta_4^2 (0)
\frac{\theta_2(u )\theta_3(u )}{\theta_1(u)\theta_2(u )}\,.
$$
From (\ref{ET2}) it follows that
$$
R=e^{S(\eta)}, \quad \phantom{a}
\frac{C}{R}=\frac{2S'(\eta)}{\pi \theta_2(0)\theta_3(0)}.
$$
Taking logarithms of equations (\ref{ET3}) and applying
$\p_{t_0}$ and $\p_{\bar t_0}$, we represent the dPfaff-Toda hierarchy in  
the following form:
\beq\label{SS1}
\begin{array}{l}
\nabla (z_1)S\Bigl (u(z_2)\Bigr )=\phantom{}
\p_{t_0}S\Bigl ( u(z_1)\! -\! u(z_2) \Bigr ),
\;\;
\nabla (z_1)S\Bigl (u(z_2)+\eta \Bigr )=\phantom{}
\p_{\bar t_0}S\Bigl ( u(z_1)\! -\! u(z_2) \Bigr ),
\\ \\
\bar \nabla (\bar z_1)S\Bigl (u(z_2)\Bigr )=\phantom{}
\p_{t_0}S\Bigl ( \bar u(\bar z_1)\! +\! u(z_2)\! +\! \eta \Bigr ),
\;\;
\bar \nabla (\bar z_1)S\Bigl (u(z_2)\! +\! \eta\Bigr )=\phantom{}
\p_{\bar t_0}S\Bigl ( \bar u(\bar z_1)\! +\! u(z_2)\! +\! \eta \Bigr ),
\end{array}
\eeq
together with complex conjugate equations.
In particular, in the limit $z_2\to \infty$ we have 
$$\nabla (z)\log R = \p_{\bar t_0}S(u(z))=
\p_{t_0}S(u(z)+\eta).$$
Equations (\ref{SS1}) is the starting point for
investigating one-variable reductions.

\section{One-variable reductions}

One may look for solutions of the dPfaff-Toda hierarchy such that 
$u(z, {\bf t})$, $\eta ({\bf t})$ and $\tau ({\bf t})$ depend on the times 
through a single variable $\lambda = \lambda ({\bf t})$:
$u(z, {\bf t})=u(z, \lambda ({\bf t}))$, $\eta ({\bf t})=\eta (\lambda ({\bf t}))$,
$\tau ({\bf t})=\tau (\lambda ({\bf t}))$. 
Such solutions are called one-variable reductions. 
Our goal is to characterize the class of functions 
$u(z, \lambda )$, $\eta (\lambda )$, $\tau (\lambda )$ that are consistent with
the hierarchy. For simplicity, in what follows we put $\lambda =\tau$.

In this section we use the notation
$$
E^{(a)}(u)=E^{(a)}(u|\tau )=\p_u \log \theta_a (u|\tau )
$$
for logarithmic derivatives of the theta functions.
For brevity we also set
$$
E(u):=E^{(1)}(u|\tau )+E^{(4)}(u|\tau )=E^{(1)}\Bigl ( u\Bigm |\frac{\tau}{2}\Bigr ).
$$

For partial $\tau$-derivatives of the $S$-function we write
$\dot S(u)=\p_{\tau}S(u|\tau )$.
In the case when the argument of the $S$-function depends on $\tau$
the full $\tau$-derivative is given by
\beq\label{SS4}
\frac{dS(u)}{d\tau}=S'(u)\p_{\tau}u +\dot S(u).
\eeq
In what follows we need the identities 
\beq\label{SS2}
4\pi i \dot S(u)=2S'(u)E^{(2)}(u)+\pi^2 \theta_4^4(0)
\eeq
and
\beq\label{SS3}
S'(x_1-x_2) \Bigl (-E(x_1)+E(x_2)+
2E^{(2)}(x_1-x_2)\Bigr )+\pi^2 \theta_4^4(0)=
S'(x_1)S'(x_2).
\eeq
These identities are proved in \cite{AZ14}. 
Plugging (\ref{SS2}) into (\ref{SS4}), we have
\beq\label{SS4a}
4\pi i \frac{dS(u)}{d\tau}=S'(u)\Bigl (4\pi i \p_{\tau}u + 2E^{(2)}(u)\Bigr )
+\pi^2 \theta_4^4(0).
\eeq

Assuming the one-variable reduction, 
one can see that
after the substitutions 
\beq
\label{L1}
\left \{ \begin{array}{rcl}4\pi i\p_{\tau}\eta &=&
E(\xi -\eta )-E(\xi )
\\ && \\
4\pi i\p_{\tau}u &=& -E(u+\xi )+E(\xi )
\\ && \\
4\pi i\p_{\tau}\bar u &=& -E(\bar u+\bar \xi )+E(\bar \xi )
\end{array}\right.
\eeq
with the condition
\beq\label{L2}
\xi +\bar \xi =\eta
\eeq
equations (\ref{SS1}) become identities (some details 
of the calculations are given in the appendix). This just means that the reduction 
is consistent, with $\xi$, $\bar \xi$ being any functions of $\tau$ 
constrained by $\xi (\tau )+\bar \xi (\tau )=\eta (\tau )$. 
We note that $-\eta$ obeys the same differential equation as $u(z)$.

Taking into account the constraint $\xi +\bar \xi =\eta$ one can set 
\beq\label{L3}
\xi (\tau )=\frac{\eta (\tau )}{2}+i\kappa (\tau ), \qquad
\bar \xi (\tau )=\frac{\eta (\tau )}{2}-i\kappa (\tau ),
\eeq
where $\kappa (\tau )$ is an arbitrary real-valued function which plays the role of 
the ``driving function''. Then the equations (\ref{L1}) acquire the form
\beq\label{L4}
\left \{ \begin{array}{rcl}4\pi i\, \p_{\tau}\eta (\tau )&=&
-E(\frac{\eta}{2}+i\kappa )-E(\frac{\eta}{2}-i\kappa )
\\ && \\
4\pi i\p_{\tau}u (z, \tau ) &=& -E(u+\frac{\eta}{2}+i\kappa )+E(\frac{\eta}{2}+i\kappa )
\\ && \\
4\pi i\p_{\tau}\bar u(z, \tau )  &=& 
-E(u+\frac{\eta}{2}-i\kappa )+E(\frac{\eta}{2}-i\kappa )
\end{array}\right.
\eeq
These equations are sufficient conditions for 
the functions $u (z, \tau )$, $\bar u(z, \tau )$ and $\eta (\tau )$ to be
compatible with the infinite dPfaff-Toda hierarchy. Given $\kappa (\tau )$,
one should solve the first equation for $\eta (\tau )$ and substitute it
to the other two equations.

Equations of the reduced hierarchy are written for the dependent variable
$\tau$. In order to obtain them, we need the relations
\beq\label{L5}
\displaystyle{\nabla (z)\tau = \frac{d\tau }{d\log \rho}\, \nabla (z) \log \rho =
\frac{dS (u(z))/d\tau}{d\log \rho/d\tau}\, \p_{t_0}\tau},
\eeq
\beq\label{L6}
\displaystyle{\bar \nabla (\bar 
z)\tau = \frac{d\tau }{d\log R}\, \bar \nabla (\bar z) \log R =
\frac{dS (\bar u(\bar z))/d\tau}{dS(\eta )/d\tau}\, \p_{t_0}\tau}
\eeq
which are easily obtained using the chain rule of differentiating.
Their right hand sides can be further transformed with the help of 
formulas (\ref{ap1}), (\ref{ap2}), (\ref{ap3}), (\ref{ap4}) from the 
appendix. As a result, we obtain
\beq\label{L7}
\nabla (z)\tau = \frac{S'(u(z)+\xi )}{S'(\xi )}\, \p_{t_0}\tau \,, 
\qquad
\bar \nabla (\bar z)\tau = -\, \frac{S'(\bar u(\bar z)+\bar \xi )}{S'(\xi )}\, \p_{t_0}\tau .
\eeq
These are the generating equations for the infinite reduced hierarchy of 
equations of hydrodynamic type. In order to write them explicitly, we 
employ the expansion
\beq\label{L8}
S(u(z)+v)=S'(v)+\sum_{k\geq 1}\frac{z^{-k}}{k}B'_k(v)\,, \quad k\geq 1.
\eeq
(The functions $B_k = B_k(v|\tau )$ are elliptic analogues of the Faber
polynomials, $B_k'(v)=\p_v B_k (v)$.) 
Then the equations are as follows:
\beq\label{L9}
\frac{\p \tau}{\p t_k}=\phi_k (\xi (\tau )|\tau )\frac{\p \tau}{\p t_0}\,,
\qquad
\frac{\p \tau}{\p \bar t_k}=\psi_k (\xi (\tau )|\tau )\frac{\p \tau}{\p t_0}\,,
\eeq
where
\beq\label{L10}
\phi_k (\xi (\tau )|\tau )=\frac{B_k'(\xi (\tau )|\tau )}{S'(\xi (\tau )|\tau )}\,,
\qquad
\psi_k (\xi (\tau )|\tau )=-\frac{\bar B_k'(\bar \xi (\tau )|\tau )}{S'(\xi (\tau )|\tau )}.
\eeq
Formally we can extend this system to the value $k=0$ by setting 
$B_0'(u)=S'(u)$. At $k=0$ we get the equation
\beq\label{L11}
S'(\bar \xi )\p_{t_0}\tau +S'(\xi )\p_{\bar t_0}\tau =0.
\eeq
The common solution to these equations can be represented in the 
hodograph form:
\beq\label{L12}
\sum_{k\geq 1}t_k \phi_k(\xi (\tau ))+\sum_{k\geq 0} \bar t_k 
\psi_k(\xi (\tau ))=\Phi (\tau ).
\eeq
Here $\Phi$ is an arbitrary function of $\tau$.

\section{Appendix}

\subsection*{Theta functions}

The Jacobi's theta functions $\theta_a (u)=
\theta_a (u|\tau )$, $a=1,2,3,4$, are defined by the formulas
\beq\label{Bp1}
\begin{array}{l}
\theta _1(u)=-\displaystyle{\sum _{k\in \z}}
\exp \left (
\pi i \tau (k+\frac{1}{2})^2 +2\pi i \phantom{}
(u+\frac{1}{2})(k+\frac{1}{2})\right ),
\\
\theta _2(u)=\displaystyle{\sum _{k\in \z}}
\exp \left (
\pi i \tau (k+\frac{1}{2})^2 +2\pi i \phantom{}
u(k+\frac{1}{2})\right ),
\\
\theta _3(u)=\displaystyle{\sum _{k\in \z}}
\exp \left ( \phantom{}
\pi i \tau k^2 +2\pi i u k \right ),
\\
\theta _4(u)=\displaystyle{\sum _{k\in \z}}
\exp \left (
\pi i \tau k^2 +2\pi i \phantom{}
(u+\frac{1}{2})k\right )
\end{array}
\eeq 
with ${\rm Im}\, \tau >0$. The function 
$\theta_1(u)$ is odd, the other three functions are even.
Shifts by the half-periods relate
the different theta functions to each other.
We also mention the identity
\beq\label{theta1prime}
\theta_1'(0)=\pi \phantom{} \theta_2(0) \theta_3(0) \theta_4(0).
\eeq
Many useful identities for the theta functions can be found 
in \cite{KZtheta}.

\subsection*{Some details of the calculations}

Let us start with the first equation in (\ref{SS1}),
$$
\nabla (z_1)S\Bigl (u(z_2)\Bigr )=
\p_{t_0}S\Bigl ( u(z_1)\! -\! u(z_2) \Bigr ).
$$
Applying the chain rule of differentiating we can write its 
left hand side as
$$
\nabla (z_1)S(u_2)=\nabla (z_1)\tau \, \frac{dS(u_2)}{d \tau}
=\nabla (z_1)\log \rho \left (\frac{d\log \rho}{d\tau }\right )^{-1}
\frac{dS(u_2)}{d \tau}
$$
(where $\rho$ is defined in (\ref{rho}) and we have put $u_i\equiv u(z_i)$ for brevity).
In its turn, $\nabla (z_1)\log \rho$ can be found from the $z_2\to \infty$ 
limit of the first equation in (\ref{SS1}):
$$
\nabla (z_1)\log \rho =\p_{t_0}S(u_1)=\p_{t_0}\tau \, 
\frac{dS(u_1)}{d \tau}.
$$
Therefore, the left hand side is
$$
\nabla (z_1)S(u_2)=\p_{t_0}\tau \, 
\frac{dS(u_1)}{d \tau}\, \frac{dS(u_2)}{d \tau}
\left (\frac{d\log \rho}{d\tau }\right )^{-1}.
$$
The right hand side of the first equation in (\ref{SS1}) reads
$$
\p_{t_0}S(u_1-u_2)=\p_{t_0}\tau \, \frac{dS(u_1-u_2)}{d \tau}.
$$
Assuming that $\p_{t_0}\tau \neq 0$ we see that the first equation 
in (\ref{SS1}) becomes
\beq\label{SS5}
\frac{dS(u_1)}{d \tau}\, \frac{dS(u_2)}{d \tau}=
\frac{d\log \rho}{d\tau }\, \frac{dS(u_1-u_2)}{d \tau}.
\eeq
Each full $\tau$-derivative here can be further transformed 
with the help of (\ref{SS4a}).
Substituting $\p_{\tau}u$ from (\ref{L1}) and
using identity (\ref{SS3}), we can prove that
\beq\label{ap1}
4\pi i \,\frac{dS(u(z))}{d \tau}=S'(u(z)+\xi )S'(\xi )
\eeq
and, in particular (at $z\to \infty$),
\beq\label{ap2}
4\pi i \, \frac{d\log \rho}{d \tau}=(S'(\xi ))^2.
\eeq
Therefore, we see that equation (\ref{SS5}) is identically satisfied:
$$
S'(u_1+\xi )S'(\xi )\cdot S'(u_2+\xi )S'(\xi )=(S'(\xi ))^2\cdot
S'(u_1+\xi )S'(u_2+\xi ).
$$

The other equations in (\ref{SS1}) can be processed in a similar way.
Consider the second equation, 
$$
\nabla (z_1)S\Bigl (u(z_2)+\eta \Bigr )=
\p_{\bar t_0}S\Bigl ( u(z_1)\! -\! u(z_2) \Bigr ).
$$
In the left hand side we have:
$$
\begin{array}{lll}
\nabla (z_1)S(u_2+\eta )&=&\displaystyle{\nabla (z_1)\tau \, 
\frac{dS(u_2+\eta )}{d\tau }}
\\ &&\\
&=&\displaystyle{\nabla (z_1)\log R \, \frac{dS(u_2+\eta )}{d\tau }
\left (\frac{d\log R}{d\tau }\right )^{-1}
}
\\ &&\\
&=&\displaystyle{\p_{\bar t_0}S(u_1)\, \frac{dS(u_2+\eta )}{d\tau }
\left (\frac{d\log R}{d\tau }\right )^{-1}
}
\\ &&\\
&=&\displaystyle{\p_{\bar t_0}\tau  \, \frac{dS(u_1)}{d\tau }\,
\frac{dS(u_2+\eta )}{d\tau }
\left (\frac{d\log R}{d\tau }\right )^{-1}.}
\end{array}
$$
The right hand side is
$$
\p_{\bar t_0}S(u_1-u_2)=\p_{\bar t_0}\tau  \,\frac{dS(u_1-u_2 )}{d\tau }.
$$
Recall also that $\log R=S(\eta )$.
Therefore, the second equation 
in (\ref{SS1}) becomes
\beq\label{SS6}
\frac{dS(u_1)}{d \tau}\, \frac{dS(u_2+\eta )}{d \tau}=
\frac{dS(\eta )}{d\tau }\, \frac{dS(u_1-u_2)}{d \tau}.
\eeq
Again, substituting $\p_{\tau}u$ from (\ref{L1}) and
using identity (\ref{SS3}), we can prove that
\beq\label{ap3}
4\pi i \,\frac{dS(u_2+\eta )}{d \tau}=S'(u_2+\xi )S'(\xi -\eta ),
\eeq
\beq\label{ap3a}
4\pi i \,\frac{dS(u_1-u_2)}{d \tau}=S'(u_1+\xi )S'(u_2+\xi ),
\eeq
and, in particular (at $z\to \infty$ in (\ref{ap3})),
\beq\label{ap4}
4\pi i \, \frac{dS(\eta )}{d \tau}=S'(\xi ) S'(\xi -\eta )
=S'(\bar \xi ) S'(\bar \xi -\eta )
\eeq
(the last equality holds because $\xi +\bar \xi =\eta$).
Therefore, we see that equation (\ref{SS6}) is identically satisfied:
$$
S'(u_1+\xi )S'(\xi )\cdot S'(u_2+\xi )S'(\xi -\eta )=S'(\xi ) 
S'(\xi -\eta  )\cdot S'(u_1+\xi )S'(u_2+\xi ).
$$

Now consider the third equation in (\ref{SS1}), 
$$
\bar \nabla (\bar z_1)S\Bigl (u(z_2)\Bigr )=
\p_{t_0}S\Bigl ( \bar u(\bar z_1)\! +\! u(z_2)\! +\! \eta \Bigr ).
$$
Its left hand side is
$$
\begin{array}{lll}
\bar \nabla (\bar z_1)S(u_2)&=&\displaystyle{\bar \nabla (\bar z_1)\tau \, 
\frac{dS(u_2)}{d\tau }}
\\ &&\\
&=&\displaystyle{\bar \nabla (\bar z_1)\log R \, \frac{dS(u_2)}{d\tau }
\left (\frac{d\log R}{d\tau }\right )^{-1}
}
\\ &&\\
&=&\displaystyle{\p_{t_0}S(\bar u_1)\, \frac{dS(u_2)}{d\tau }
\left (\frac{d\log R}{d\tau }\right )^{-1}
}
\\ &&\\
&=&\displaystyle{\p_{t_0}\tau  \, \frac{dS(\bar u_1)}{d\tau }\,
\frac{dS(u_2)}{d\tau }
\left (\frac{d\log R}{d\tau }\right )^{-1}.}
\end{array}
$$
When passing from the second line to the third one we have used 
the complex conjugate of the second equation in (\ref{SS1}) 
in the limit $z_2\to \infty$. The right hand side is
$$
\p_{t_0}S(\bar u_1+u_2+\eta )=\p_{t_0}\tau  
\,\frac{dS(\bar u_1+u_2+\eta)}{d\tau }.
$$
The third equation 
in (\ref{SS1}) becomes
\beq\label{SS7}
\frac{dS(\bar u_1)}{d \tau}\, \frac{dS(u_2)}{d \tau}=
\frac{dS(\eta )}{d\tau }\, \frac{dS(\bar u_1+u_2+\eta )}{d \tau}.
\eeq
Substituting here $\p_{\tau}u$ from (\ref{L1}) and
using identity (\ref{SS3}), we can prove that
\beq\label{ap5}
4\pi i \,\frac{dS(\bar u(\bar z))}{d \tau}=S'(\bar u(\bar z)+\bar \xi )S'(\bar \xi )
\eeq
and 
\beq\label{ap6}
4\pi i \,\frac{dS(\bar u_1 +u_2 +\eta )}{d \tau}=
S'(\bar u_1 +\bar \xi )S'(-u_2 -\xi ).
\eeq
Therefore, we see that equation (\ref{SS7}) is identically satisfied:
$$
S'(\bar u_1+\bar \xi )S'(\bar \xi )\cdot S'(u_2+\xi )S'(\xi )=S'(\xi ) 
S'(\xi -\eta  )\cdot S'(\bar u_1+\bar \xi )S'(-u_2-\xi )
$$
(we use the fact that $S'(u)$ is odd function and the constraint 
$\xi +\bar \xi =\eta$).

The calculations for the remaining case of the fourth equation 
in (\ref{SS1}) are similar.

\section*{Acknowledgements}


We thank T.Takebe for discussions.
This work has been funded by the  Russian Academic Excellence Project ``5-100''. 
Results of section 3 has been obtained under support of the RSF grant 16-11-10160.
Research of both authors has also been supported by RFBR grant
14-02-00627. The work of A.Z. has also been partially supported 
by joint RFBR grant 15-52-50041-YaF.

\end{document}